# Scale-Dependent Pedotransfer Functions Reliability for Estimating Saturated Hydraulic Conductivity

Behzad Ghanbarian[1*], Vahid Taslimitehrani[2] and Yakov A. Pachepsky[3]

[1] Bureau of Economic Geology, Jackson School of Geosciences, University of Texas at Austin, Austin, TX 78713

[2] PhysioSigns Inc., Sunnyvale, CA 94086

[3] Environmental Microbial Safety Lab., USDA-ARS, Beltsville, MD 20705

[*] Corresponding author's email address: behzad.ghanbarian@beg.utexas.edu

**Abstract**

Saturated hydraulic conductivity $K_{sat}$ is a fundamental characteristic in modeling flow and contaminant transport in soils and sediments. Therefore, many models have been developed to estimate $K_{sat}$ from easily measureable parameters, such as textural properties, bulk density, etc. However, $K_{sat}$ is not only affected by textural and structural characteristics, but also by sample dimensions e.g., internal diameter and height. Using the UNSODA database and the contrast pattern aided regression (CPXR) method, we recently developed sample dimension-dependent pedotransfer functions to estimate $K_{sat}$ from textural data, bulk density, and sample dimensions. The main objectives of this study were evaluating the proposed pedotransfer functions using a larger database, and comparing them with seven other models. For this purpose, we selected more than nineteen thousands soil samples from all around the United States. Results showed that the sample dimension-dependent pedotransfer functions estimated $K_{sat}$ more accurately than seven other models frequently used in the literature.

**Keywords:** Contrast pattern aided regression, Pedotransfer function, Sample dimensions, Saturated hydraulic conductivity

## 1. Introduction

Saturated hydraulic conductivity $K_{sat}$ is essential in modeling surface and subsurface flow as well as solute transport in soils and sediments. Its estimation has been intensively investigated over the past several decades by soil physicists and hydrologists. Many empirical (e.g., Hazen, 1892; Puckett et al., 1985; Nemes et al., 2005; Parasuraman et al., 2006; Ghanbarian-Alavijeh et al., 2010), quasi-physical (e.g., Kozeny, 1927; Carman, 1937; Ahuja et al., 1984;1989; Rawls et al. 1993; Arya et al. 1999; Timlin et al. 1999), and physically-based (e.g., Katz and Thompson 1986; Xu and Yu 2008; Skaggs, 2011; Porter et al. 2013; Hunt et al. 2014; Ghanbarian et al. 2016) models as well as numerical techniques (see e.g., Zhang et al. 2005; Elliot et al. 2010; Mostaghimi et al. 2013; Ghanbarian and Daigle 2015; Dal Ferro and Morari 2015) were developed to estimate $K_{sat}$ from other available properties. Among them, pedotransfer functions i.e., regression-based models making use of available information to provide estimates of other related factors that are needed (Warrick, 2003), attracted a great deal of attention mainly due to their relatively high efficiency and simplicity. Pedotransfer functions also have limitations. For example, Schaap and Leij (1998a) showed that the performance of pedotransfer functions may strongly depend on data that were used to train (or develop) and test (or validate). Nonetheless, pedotransfer functions have been actively developed to link soil hydraulic properties to readily measureable characteristics e.g., soil texture, bulk density, and organic matter content.

To estimate the saturated hydraulic conductivity from other soil parameters pedotransfer functions have been developed using various techniques, such as multiple linear regression (e.g., Brakensiek et al., 1984; Cosby et al., 1984; Saxton et al., 1986; Jabro 1992), artificial neural networks (e.g., Schaap and Leij, 1998b; Schaap et al., 1998; 2001; Parasuraman et al., 2006; Ghanbarian-Alavijeh et al., 2010), genetic programming (e.g., Parasuraman et al., 2007), decision tree analysis (e.g., Shirley, 2011; Jorda et al., 2015), group method of data handling (e.g., Nemes et al., 2005), support vector machines (e.g., Twarakavi et al., 2009), and contrast pattern aided regression (Ghanbarian et al., 2015).

Minasny and McBratney (2000) evaluated eight proposed pedotransfer functions for estimation of saturated hydraulic conductivity $K_{sat}$ using published Australian soil data sets comprised different field and laboratory measurements over large areas and limited predictive variables. They found that the published PTFs of Dane and Puckett (1994), Cosby et al. (1984), and Schaap et al. (1998) gave the best estimations for sandy, loamy, and clayey soils, respectively.

Sobieraj et al. (2001) evaluated the performance of nine $K_{sat}$ pedotransfer functions including the Rosetta model (see their Table 3) in modeling the storm flow generated in a rainforest catchment and found them inaccurate. At shallow depths e.g., 0-0.1 m, pedotransfer functions, except Jabro (1992), commonly underestimated $K_{sat}$, while at subsequent depths e.g., 0.1-0.4 m pedotransfer functions, except Campbell and Shiozawa (1994), typically overestimated $K_{sat}$.

Gwenzi et al. (2011) evaluated five PTFs including Cosby et al. (1984), Jabro (1992), Puckett et al. (1985), Dane and Puckett (1994), and Saxton et al. (1986). They showed that all evaluated $K_{sat}$ models underestimated by an order of magnitude, suggesting that application of water balance simulation models based on such PTFs may constitute a bias in model outputs.

More recently, Ghanbarian et al. (2015) proposed sample dimension-dependent pedotransfer functions to estimate saturated hydraulic conductivity $K_{sat}$ using the contrast pattern aided regression (CPXR) method and soil samples from the UNSODA database. They showed that including sample dimensions i.e., sample internal diameter and height (or length) substantially improved $K_{sat}$ estimations.

Although numerous studies assessed the performance of different pedotransfer functions in estimation of water content and soil water retention curve, such evaluations for $K_{sat}$ pedotransfer functions and estimations are limited in the literature and particularly restricted to assessments at small scales. Therefore, the main objective of this study was to use a large database including soil samples from all around the United States for (1) evaluating the sample dimension-dependent pedotransfer functions developed by Ghanbarian et al. (2015) using the contrast pattern aided regression (CPXR) method in estimation of saturated hydraulic conductivity $K_{sat}$ from clay, silt, and sand contents, geometric mean diameter and standard deviation of diameters of soil particles, bulk density, and sample dimensions (i.e., internal diameter and height), and (2) comparing performance of the proposed CPXR-based pedotransfer functions with performance of seven other pedotransfer functions frequently applied in the literature.

## 2. Materials and Methods

Here, we first describe datasets used in this study and then present proposed pedotransfer functions available in the literature for estimating saturated hydraulic conductivity from clay, silt, and sand content as well as bulk density.

### 2.1. Datasets

The entire database used in this study includes 24 datasets and 19822 soil samples selected from the USKSAT database (Pachepsky and Park, 2015) and different states in the nation. The number of samples, $K_{sat}$ measurement method, and average and standard deviation of different parameters for each dataset are summarized in Table 1. The average $K_{sat}$ value varying between 12.3 (Jabro, 1992) and 1328.5 cm/day (Romkens et al., 1986), near three orders of magnitude variation, indicates different types of soils with various structures. This is an order of magnitude greater than the average $K_{sat}$ range that Schaap et al. (1998) investigated through 620 soil samples from 30 sources in the United States. In the Results and Discussion section below, we however, show that the measured $K_{sat}$ value actually spans even much more than 3 orders of magnitude (about 8) representing a very wide range of soils. Figure 1 showing the percentage of samples in each USDA soil texture class and the distribution of samples on the ternary diagram also implies various types of soils with different textures used in this study.

### 2.2. Saturated hydraulic conductivity pedotransfer functions

There are tremendous amount of models in the literature estimating the saturated hydraulic conductivity from other properties. However, the focus of our study is $K_{sat}$ estimation from frequently available characteristics, such as soil textural data i.e., sand, silt, and clay content and bulk density for soil mapping purposes at large scales. In Table 2, we list 7 models commonly used in the literature including Brakensiek et al. (1984), Campbell and Shiozawa (1994), Cosby et al. (1984), Jabro (1992), Puckett et al. (1985), Dane and Puckett (1994), and Saxton et al. (1986).

Both Campbell and Shiozawa (1994) and Puckett et al. (1985) used the same database consisting of 42 samples (six Ultisols taken at seven locations in the lower coastal Plain of Alabama). The Cosby et al. (1984) model was proposed based on 1448 samples from Holtan et al. (1968) and Rawls et al. (1976). Jabro (1992) presented his model using 350 soil samples collected from the literature e.g., Southern Cooperation Series Bulletins. Dane and Puckett (1994) developed their model using 577 samples (Ultisols from lower coastal Plain of Alabama). The Saxton et al. (1986) model was developed using a database including 230 selected data points uniformly spaced on the hydraulic conductivity curves for 10 textures classes reported by Rawls et al. (1982).

All these models, except Brakensiek et al. (1984), estimate $K_{sat}$ from sand, silt, and clay content and bulk density $\rho_b$. The Brakensiek et al. (1984) model, however, requires porosity value as well, which was estimated by $1 - \rho_b/\rho_s$ in which $\rho_s$, the particle density, was assumed to be approximately equal to 2.65 g/cm$^3$ for all samples.

We should point out that we eliminated the Jabro (1992) dataset (see Table 1) from this model evaluation process since the assessment of the Jabro (1992) model using the same dataset used to develop it would be biased and not supported. In addition, the Jabro (1992) model estimates $K_{sat}$ from logarithms of clay and silt content (see Table 2). As a consequence, the Jabro (1992) model returns unrealistic $K_{sat}$ estimate for samples with either zero clay or silt content. Removing the Jabro (1992) dataset as well as those samples with either zero clay or silt content resulted in $n = 19480$ in evaluation of the Jabro (1992) model.

The sample dimension-dependent model of Ghanbarian et al. (2015) presented in Table 3 consists of 14 patterns each of which corresponds to a specific model. Most patterns given in Table 3 match more than 10% (i.e., support > 10%) of the training data. As Ghanbarian et al. (2015) noted, even patterns with supports less than that still match about 10 samples of the training data. Moreover, patterns having relatively small supports often have large ARR (average residual reduction) values, implying that the matching samples of each such pattern represent an important data group where the baseline model $f_0$ makes very large prediction errors and those errors are reduced significantly by the local model (see Table 3) corresponding to each pattern (Ghanbarian et al., 2015).

To estimate $K_{sat}$ using the Ghanbarian et al. (2015) model, first the relevant pattern that the data match with is found, and then using the corresponding local model given in Table 3 $K_{sat}$ is estimated. One should expect that one soil sample may match more than one pattern. In that case, all relevant patterns should be identified and used to estimate $K_{sat}$. The weighted average of all $K_{sat}$ values estimated by different patterns and local models then represents the Ghanbarian et al. (2015) model estimation. In case a sample does not match any pattern, $K_{sat}$ is estimated using the baseline model $f_0$ given in Table 3. The detail of the CPXR approach can be found in Ghanbarian et al. (2015). The Java code for $K_{sat}$ estimation by the Ghanbarian et al. (2015) model and other models presented in Table 2 is given in the Supplementary Material.

In order to compare statistically the accuracy of the discussed models in the estimation of $K_{sat}$, the root mean square log-transformed error (RMSLE) and mean log-transformed error (MLE) parameters were determined as follows

$$RMSLE = \sqrt{\frac{1}{n}\sum_{i=1}^{n}[\log(K_{sat})_{cal} - \log(K_{sat})_{meas}]^2} \qquad (1)$$

$$MLE = \frac{1}{n}\sum_{i=1}^{n}[\log(K_{sat})_{cal} - \log(K_{sat})_{meas}] \qquad (2)$$

where $n$ is the number of values, and $(K_{sat})_{cal}$ and $(K_{sat})_{meas}$ are the calculated (estimated) and measured saturated hydraulic conductivities, respectively.

## 3. Results and Discussion

Comparison of the performance of the selected models in the estimation of saturated hydraulic conductivity $K_{sat}$ from sand, silt and clay content as well as bulk density is shown in Fig. 2. As can be seen, the Campbell and Shiozawa (1994) and Puckett et al. (1985) models with MLE = -0.96 and -0.41 tend to underestimate $K_{sat}$ considerably over 8 orders of magnitude variation, while the Dane and Puckett (1994), Saxton et al. (1986) and Ghanbarian et al. (2015) models estimates are distributed more or less around the 1:1 line (MLE = 0.09, -0.06, and 0.08, respectively). Interestingly, although the Puckett et al. (1985) and Campbell and Shiozawa (1994) models were developed using the same exponential function and dataset (42 samples from the lower coastal Plain of Alabama) and have similar trend in $K_{sat}$ underestimation, the Puckett et al. (1985) model is almost 25% more accurate than the Campbell and Shiozawa (1994) model (RMSLE = 0.92 vs. 1.27). Such a discrepancy in $K_{sat}$ estimation implies that the Puckett et al. (1985) model could more efficiently detect nonlinear interactions between input and output variables than the Campbell and Shiozawa (1994) model. The MLE values for different $K_{sat}$ models and USDA soil texture classes presented in Table 4 also show that both Puckett et al. (1985) and Campbell and Shiozawa (1994) tend to underestimate $K_{sat}$ for all soil texture classes, particularly intermediate to fine textured soils, while the Jabro (1992) model has a tendency to overestimate $K_{sat}$.

As can be observed in Fig. 2, although the value of mean log-transformed error of the Cosby et al. (1984) model is near zero (MLE = -0.001), the $K_{sat}$ estimations are biased since the model overestimates at lower $K_{sat}$ (fine-textured soils) and underestimates at higher $K_{sat}$ values (coarse-textured soils). In this model, the measured $K_{sat}$ spans around 8 orders of magnitude ($\log(K_{sat})$ varies between -2 to near 6), while the estimated $\log(K_{sat})$ value is restricted to a much narrower range approximately between 0 and 3. The obtained results are, however, consistent with the results of Minasny and McBratney (2000) who found that the Cosby et al. (1984) model gave the best $K_{sat}$ estimations for loamy soils with intermediate $K_{sat}$. Using 462 soil samples from Australia, Minasny and McBratney (2000) reported RMSLE = 6.41, 2.58, 2.71, 3.08, and 3.53 [ln(mm/hr)] for the Campbell and Shiozawa (1994), Cosby et al. (1984), Dane and Puckett (1994), Brakensiek et al. (1984), and Saxton et al. (1986) models, respectively, meaning the Cosby et al. (1984) model was the most accurate. These RMSLE values, however, are not comparable with those we present in Fig. 2 because (1) the $K_{sat}$ units is different (mm/hr vs. cm/day), (2) the statistical parameter calculated in Minasny and McBratney (2000) is the root mean square *natural* log-transformed error, while that we report is root mean square log-transformed error, and (3) the number of samples in Minasny and McBratney (2000) is 462, whereas *n* in our study is 19822, resulted into RMSLE values much smaller than those reported by Minasny and McBratney (2000).

Figure 2 also shows that the Jabro (1992) model tends to overestimate $K_{sat}$ substantially (MLE = 0.57). Although our results obtained at large scale might be in contrast with the results of Gwenzi et al. (2011) who reported that the Jabro (1992) model underestimated $K_{sat}$ by an order of magnitude, the number of samples in our study compared to that in Gwenzi et al. (2011) is much larger and one thus expect the obtained results to be more general.

The RMSLE values for different $K_{sat}$ models and USDA soil texture classes are given in Table 5. As the bold RMSLE values in Table 5 indicate the most accurate model estimating $K_{sat}$ for different soil texture classes precisely is Ghanbarian et al. (2015). Among all the models evaluated in this study and shown in Fig. 2, the estimations of the Ghanbarian et al. (2015) model are more appropriately and uniformly distributed around the 1:1 line indicating the most accurate model. Comparison of the RMSLE parameter calculated for each model shows that the value of RMSLE = 0.56 for the Ghanbarian et al. (2015) model is considerably smaller than that determined for other models. After Ghanbarian et al. (2015), the Dane and Puckett (1994) model estimates $K_{sat}$ slightly better than the Saxton et al. (1986) and Cosby et al. (1984) models. Interestingly, both Dane and Puckett (1994) and Puckett et al. (1985) models have the same exponential function and input variable i.e., clay. However, the Dane and Puckett (1994) model with RMSLE = 0.73 estimated $K_{sat}$ more accurately than the Puckett et al. (1985) model with RMSLE = 0.92. The main reason probably is in the former 577 soil samples were used to construct a relationship between $K_{sat}$ and clay content, while in the latter only 42 samples. As a consequence, the larger the training database, the more reliable the validation results are. Figure 2 also shows that the Brakensiek et al. (1984) model estimated $K_{sat}$ more accurately than the Puckett et al. (1985) and Campbell and Shiozawa (1994) models. One should note that the calculated RMSLE value for the Jabro (1992) model is not directly comparable with those determined for other models since this model is using logarithms of clay and silt contents and, therefore, the number of samples suitable to evaluate this model is 342 less than that used to assess others. Results of our study confirm that the sample dimension-dependent Ghanbarian et al. (2015) model is the most accurate in the estimation of the soil saturated hydraulic conductivity $K_{sat}$. The main reasons for that are discussed below.

First, the model proposed by Ghanbarian et al. (2015) is sample dimension-dependent, which means it requires more input variables such as sample height (or length) and internal diameter to estimate $K_{sat}$. The influence of sample dimension on $K_{sat}$ has been well documented in the literature. For example, recently Pachepsky et al. (2014) showed the similarity of scale (sample dimension) dependencies in soils and sediments. They observed that as the characteristic size increased, $K_{sat}$ value first increased by one or two orders of magnitude and then stabilized. More recently, Ghanbarian et al. (2015) showed that the influence of sample dimensions (internal diameter and height) and soil water retention curve on the estimation of the saturated hydraulic conductivity was comparable in both CPXR and multiple linear regression (MLR) methods (for more evidence comprehensive results see Ghanbarian et al. (2015) and references therein).

Second, the Ghanbarian et al. (2015) model was developed using a novel and modern data mining technique, introduced recently by Dong and Taslimitehrani (Taslimitehrani and Dong, 2014; Dong and Taslimitehrani, 2015) and called contrast pattern aided regression (CPXR), while all other models were proposed using simple algorithms such as multiple linear regressions (MLRs). Ghanbarian et al. (2015) compared CPXR-based pedotransfer functions with MLR-based ones in estimation of $K_{sat}$ and demonstrated that under all circumstances the CPXR method performed batter than the MLR technique. The main advantage of the CPXR method compared to multiple linear regression-based models is that CPXR is capable to detect complex and nonlinear interactions between input variables that define different subgroups of data with highly distinct predictor–response relationships, and to extract nonlinear patterns among input and output variables. One more advantage of the CPXR technique is that the model can be reported in the readable and transparent form (Table 3) and, unlike results of artificial neural networks or support vector machine application, can be visually inspected and analyzed.

**5. Conclusion**

In this study, we evaluated various models available in the literature to estimate soil saturated hydraulic conductivity $K_{sat}$. Using a large database including 19822 soil samples from all round the United States we compared the reliability of Brakensiek et al. (1984), Campbell and Shiozawa (1994), Cosby et al. (1984), Jabro (1992), Puckett et al. (1985), Dane and Puckett (1994), and Saxton et al. (1986) models. In addition to these models, the Ghanbarian et al. (2015) model estimating $K_{sat}$ from textural data, bulk density as well as sample dimensions e.g., height (or length) and internal diameter was also assessed. As we emphasized before, the assessment of models estimating $K_{sat}$ from other soil properties is limited in the literature. In addition, the existing evaluations are mainly restricted to fairly small databases and small scales, and therefore the obtained results reported in the literature might be in contradiction. Our results based on 19822 measurements including a very wide range of soils from all around the nation are unique and general. We conclusively showed that the proposed sample dimension-dependent pedotransfer functions of Ghanbarian et al. (2015) estimated $K_{sat}$ substantially more accurately than seven other frequently used previously published models.

**Acknowledgment**

Hubbard, R.K., C. R. Berdanier, H. F. Perkins, and R. A. Leonard. 1985. Characteristics of Selected Upland Soils of the Georgia Coastal Plain. USDA-ARS. ARS-37. U.S. Government Printing Office, Washington, DC.

Hunt, A. G., R. P. Ewing, and B. Ghanbarian (2014), Percolation Theory for Flow in Porous Media, Lecture Notes Phys., vol. 880, 3rd ed., Springer, Heidelberg, Germany.

Jabro, J. D. 1992. Estimation of saturated hydraulic conductivity of soils from particle size distribution and bulk density data. Am. Soc. Agric. Eng. 35: 557–560.

Jorda, H., Bechtold, M., Jarvis, N., & Koestel, J. (2015). Using boosted regression trees to explore key factors controlling saturated and near-saturated hydraulic conductivity. European Journal of Soil Science, 66(4), 744-756.

Katz, A. J., & Thompson, A. H. (1986). Quantitative prediction of permeability in porous rock. Physical review B, 34(11), 8179-8181.

Kool, J.B., K. A. Albrecht, J. C. Parker, J. C. Baker. 1954. Physical & Chemical characterization of the Groseclose soil mapping unit. Blacksburg, VA, Virginia Agric. Exp. Sta., Virginia Polytechnic Institute and State University. Bulletin 86-4

Kozeny, J. 1927. Über Kapillare Leitung des Wassers im Boden. Sitzungsber. Akad. Wiss. Wien 136:271–306.

Minasny, B., & McBratney, A. B. (2000). Evaluation and development of hydraulic conductivity pedotransfer functions for Australian soil. Soil Research, 38(4), 905-926.

Mostaghimi, P., Blunt, M. J., & Bijeljic, B. (2013). Computations of absolute permeability on micro-CT images. Mathematical Geosciences, 45(1), 103-125.
18

Table 1. The average and standard deviation (in parentheses) of salient properties of soil samples ($n = 19822$) from various references in the USKSAT database used in this study.

| Reference | Samples no. | Measurement method | Bulk density | Sand% | Silt% | Clay% | Sample height (cm) | Sample diameter (cm) | $K_{sat}$ (cm/day) |
|---|---|---|---|---|---|---|---|---|---|
| Kool et al. (1954) | 74 | Falling head | 1.41 (0.14) | 17.5 (8.1) | 53.7 (11.1) | 29.1 (16.4) | 4 (0) | 5.4 (0) | 110.5 (207.0) |
| Hubbard et al. (1985) | 41 | Constant head | 1.71 (0.16) | 67.8 (12.4) | 8.4 (3.1) | 23.8 (11.8) | 7.5 (0) | 7.6 (0) | 89.3 (159.8) |
| Boul and Southard (1988) | 27 | Constant head | 1.55 (0.08) | 44.8 (8.5) | 25.3 (9.4) | 29.8 (8.0) | 7.6 (0) | 7.6 (0) | 46.5 (106.7) |
| Robbins (1977) | 15 | Hydraulic head | 1.33 (0.06) | 25.2 (6.7) | 59.9 (5.0) | 14.9 (3.2) | 51 (0) | 5 (0) | 42.3 (12.2) |
| Bruce et al. (1983) | 73 | Constant head | 1.54 (0.11) | 46.8 (13.5) | 20.8 (4.3) | 32.4 (13.7) | 6 (0) | 5.4 (0) | 145.9 (224.1) |
| Peele et al. (1970) | 257 | Constant head | 1.47 (0.18) | 52.2 (22.1) | 23.5 (12.5) | 24.3 (15.3) | 9.4 (0) | 9.74 (0) | 325.7 (461.5) |
| Dane et al. (1983) | 120 | Constant head | 1.54 (0.09) | 91.9 (3.9) | 4.6 (5.2) | 3.8 (1.8) | 6.4 (1.9) | 6.8 (1.1) | 1149.0 (889.3) |
| Romkens et al. (1986) | 72 | Constant head | 1.46 (0.10) | 12.9 (9.2) | 64.6 (9.4) | 22.5 (7.2) | 5.9 (1.1) | 7 (0.83) | 1328.5 (4583.5) |
| Bathke and Cassel (1991) | 25 | Constant-head permeameter | 1.36 (0.16) | 44.9 (17.9) | 20.9 (10.0) | 28.4 (14.7) | 7.6 (0) | 7.6 (0) | 421.7 (1183.3) |
| Dane (1980) | 6 | Pressure head | 1.57 (0.09) | 83.6 (1.8) | 13.2 (1.0) | 3.2 (1.7) | 6 (0) | 5.4 (0) | 235.9 (166.6) |
| Zobeck et al. (1985) | 6 | Miscellaneous[*] | 1.49 (0.04) | 14.8 (1.3) | 44.7 (8.3) | 40.5 (7.0) | 10.5 (8.2) | 12.2 (10.0) | 129.4 (171.1) |
| Jabro (1992) | 18 | Constant head | 1.53 (0.07) | 17.3 (6.7) | 56.9 (8.5) | 25.8 (6.1) | 7.6 (0) | 7.6 (0) | 12.3 (8.4) |
| Bruce (1972) | 3 | Constant head | 1.60 (0.05) | 56.7 (18.9) | 20.0 (3.6) | 23.3 (17.2) | 6.2 (0) | 6.0 (0) | 841.7 (718.1) |
| Habecker et al. (1990) | 14 | Falling-head permeameter | 1.45 (0.32) | 35.8 (21.1) | 54.2 (21.4) | 10.0 (2.1) | 7.5 (0) | 7.6 (0) | 277.1 (522.9) |
| Radcliffe et al. (1990) | 9 | Constant head | 1.52 (0.09) | 52.0 (12.9) | 19.3 (5.1) | 28.7 (16.0) | 7.6 (0) | 7.6 (0) | 209.3 (198.4) |
| Coelho (1974) | 175 | Falling head | 1.45 (0.16) | 39.4 (21.8) | 35.4 (14.4) | 24.5 (11.1) | 7.6 (0) | 7.6 (0) | 90.6 (186.5) |
| Southard and Boul (1988) | 27 | Constant head | 1.55 (0.08) | 44.5 (8.4) | 25.4 (9.2) | 30.1 (7.4) | 7.6 (0) | 7.6 (0) | 13.9 (23.3) |
| Rawls et al. (1998) | 166 | Constant Head | 1.52 (0.13) | 90.3 (9.7) | 3.4 (3.3) | 6.3 (8.6) | 5.4 (0) | 3.0 (0) | 640.1 (737.2) |
| Carlisle et al. (1978) | 780 | Constant Head | 1.50 (0.25) | 83.7 (16.7) | 5.7 (5.9) | 10.4 (12.8) | 5.4 (0) | 3.0 (0) | 548.5 (823.1) |
| Carlisle et al. (1981) | 17869 | Constant Head | 1.53 (0.17) | 85.8 (15.9) | 5.6 (7.3) | 8.7 (11.9) | 5.4 (0) | 3.0 (0) | 548.6 (725.6) |
| Baumhardt et al. (1995) | 4 | Constant Head | 1.42 (0.17) | 40.4 (24.0) | 22.9 (16.3) | 36.8 (11.9) | 50 (0) | 75 (0) | 101.6 (75.3) |
| Romkens et al. (1985) | 36 | Constant Head | 1.41 (0.05) | 12.8 (10.5) | 69.2 (9.5) | 18.1 (8.4) | 7.5 (0) | 7.5 (0) | 21.3 (37.3) |
| Sharratt (1990) | 14 | Unknown | 1.30 (0.23) | 36.5 (25.0) | 56.7 (23.1) | 6.8 (3.9) | 2.0 (0) | 5.0 (0) | 334.9 (1009.8) |
| Price et al. (2010) | 9 | Constant Head | 1.12 (0.24) | 56.1 (4.0) | 28.0 (2.6) | 15.6 (1.7) | 7.5 (0) | 7.5 (0) | 62.4 (66.6) |

[*] Including constant head blocks, constant head cores, and falling head permeameter.

Table 2. Some proposed pedotransfer functions frequently used in the literature to estimate soil saturated hydraulic conductivity $K_{sat}$.

| Reference | $K_{sat}$ (cm/day) model* |
|---|---|
| Brakensiek et al. (1984) | $K_{sat} = 24\exp[19.52348\phi − 8.96847 − 0.028212Cl + 0.00018107Sa^2 − 0.0094125Cl^2 − 8.395215\phi^2 + 0.077718\phi Sa − 0.00298\phi^2 Sa^2 − 0.019492\phi^2 Cl^2 + 0.0000173Sa^2 Cl + 0.02733\phi Cl^2 + 0.001434\phi Sa^2 − 0.0000035Cl^2 Sa]$ |
| Campbell and Shiozawa (1994) | $K_{sat} = 129.6\exp(-0.07Si – 0.167Cl)$ |
| Cosby et al. (1984) | $K_{sat} = 60.96 \times 10^{(-0.6 + 0.0126Sa – 0.0064Cl)}$ |
| Jabro (1992) | $K_{sat} = 24 \times [9.56 – 0.81\log(Si) – 1.09\log(Cl) – 4.64\rho_b]$ |
| Puckett et al. (1985) | $K_{sat} = 376.7\exp(-0.1975Cl)$ |
| Dane and Puckett (1994) | $K_{sat} = 729.22\exp(-0.144Cl)$ |
| Saxton et al. (1986) | $K_{sat} = 24\exp(12.012 – 0.0755Sa)$ |

* Sa: sand (%), Si: silt (%), Cl: clay (%), $\rho_b$: bulk density (g cm$^{-3}$), and $\phi$: porosity.

Table 3. Patterns, criteria, baseline regression model $f_0$, and pedotransfer functions for the CPXR model developed by Ghanbarian et al. (2015) to estimate soil saturated hydraulic conductivity from sand, silt, and clay content, geometric mean and geometric standard deviation of particles, bulk density, and sample dimensions i.e., length and internal diameter. See the context for more detail about $K_{sat}$ estimation using the baseline, criteria, and local models.

| Log($K_{sat}$) [cm/day] | | | |
|---|---|---|---|
| Baseline model $f_0$: −1180 + 11.89×Sa + 11.90×Si + 11.85×Cl + 5.25×$d_g$ + 0.028×$\sigma_g$ − 3.86×$\rho_b$ − 0.039×ID | | | |
| Pattern | Criteria | ARR | Support (%) |
| 1 | 0.495 ≤ $d_g$ < 0.74, Cl < 15.8, 1.55 ≤ $\sigma_g$ < 6.96, 0.2 ≤ Si < 20. 3, 74.4 ≤ Sa < 99.1, 1.23 ≤ $\rho_b$ <1.6 | 1.5 | 16.0 |
| 2 | 49.6 ≤ Sa < 74.4, 2.5 ≤ $L$ < 36.9, 3.2 ≤ ID < 7.5 | 2.3 | 14.6 |
| 3 | 40.5 ≤ Si < 60.6, 0.004 ≤ $d_g$ < 0.25, 6.96 ≤ $\sigma_g$ < 12.38, 1.23 ≤$\rho_b$ < 1.6 | 3.1 | 8.0 |
| 4 | 0.2 ≤ Si < 20.3, Cl < 15.8, 1.55 ≤ $\sigma_g$ < 6.96, 74.4 ≤ Sa < 99.1 | 0.9 | 32.4 |
| 5 | 0.2 ≤ Si < 20.3, 0.004 ≤ $d_g$ < 0.25, 2.5 ≤ $L$ < 36.9 | 6.6 | 5.2 |
| 6 | 6.96 ≤ $\sigma_g$ < 12.38, Cl < 15.8, 2.5 ≤ $L$ < 36.9, 3.2 ≤ ID < 7.5, 20. 3 ≤ Si < 40.5, 49.6 ≤ Sa < 74.4, 0.004 ≤ $d_g$ < 0.25 | 1.6 | 6.1 |
| 7 | 60.6 ≤ Si < 80.7, 2.5 ≤ $L$ <36.9, 1.23 ≤ $\rho_b$ < 1.6, 0.004 ≤ dg < 0.25, 0.1 ≤ Sa < 24.9 | 3.4 | 14.1 |
| 8 | 49.6 ≤ Sa < 74.4, 2.5 ≤ $L$ < 36.9, 20.3 ≤ Si < 40.5, 3.2 ≤ ID < 7.5 | 2.1 | 12.7 |
| 9 | 15.8 ≤ Cl < 31.5, 2.5 ≤ $L$ <36.9, 0.004 ≤ $d_g$ < 0.25, 1.23 ≤ $\rho_b$ < 1.6 | 2.4 | 14.6 |
| 10 | 49.6 ≤ Sa < 74.4, 2.5 ≤ $L$ < 36.9, 1.23 ≤ $\rho_b$ < 1.6, 3.2 ≤ ID < 7.5 | 4.3 | 8.0 |
| 11 | 20.3 ≤ Si < 40.5, 2.5 ≤ $L$ < 36.9, Cl < 15.8, 49.6 ≤ Sa < 74.4 | 0.9 | 15.0 |
| 12 | 1.55 ≤ $\sigma_g$ < 6.96, Cl < 15.8 | 0.9 | 36.6 |
| 13 | Cl < 15.8, 0.004 ≤ $d_g$ < 0.25, 2.5 ≤ $L$ < 36.9 | 0.5 | 25.4 |
| 14 | Cl <15.8, 2.5 ≤ $L$ < 36.9 | 0.4 | 59.2 |
| Pattern | Local model corresponding to each pattern | | |
| 1 | −1355.1 + 14.07×Sa + 9.91×Si − 50.2×dg + 18.39×$\sigma_g$ − 12.99×$\rho_b$ − 0.24×ID − 0.014×$L$ | | |
| 2 | 70.74 − 0.52×Sa − 0.64×Si − 30.4×$d_g$ − 1.73×$\sigma_g$ − 4.66×$\rho_b$ + 2.67×ID − 0.025×$L$ | | |
| 3 | 52.1 + 0.58×Sa − 0.56×Si − 230.6×dg − 2.68×$\sigma_g$ − 1.18×$\rho_b$ + 0.35×ID − 0.064×$L$ | | |
| 4 | −1033.7 + 10.4×Sa + 9.7×Si + 7.6×Cl + 4.56×dg + 4.54×$\sigma_g$ − 5.14×$\rho_b$ − 0.2×ID − 0.0006×$L$ | | |
| 5 | 53.5 + 0.095×Sa − 0.54×Si − 73.4×$d_g$ − 1.58×$\sigma_g$ − 7.77× $\rho_b$ − 0.36×ID + 0.42×$L$ | | |
| 6 | 762.4 − 2.28×Sa − 7.1×Si − 770.8×dg − 22.7×$\sigma_g$ − 2.07×$\rho_b$ − 3.88×ID − 1.72×$L$ | | |
| 7 | −50.5 − 1.59×Sa + 0.56×Si + 587.4×$d_g$ + 7.4×$\sigma_g$ − 20.55×$\rho_b$ + 0.43×ID − 0.18×$L$ | | |
| 8 | −243.4 + 1.07×Sa + 2.16×Si + 156.74×dg + 5.04×$\sigma_g$ − 1.42×$\rho_b$ + 6.12×ID + 1.14×$L$ | | |
| 9 | 13.7 + 0.042×Sa − 0.104×Si − 60.1×dg − 0.035×$\sigma_g$ − 2.84×$\rho_b$ + 0.16×ID − 0.011×$L$ | | |
| 10 | 281.9 − 1.395×Sa − 2.37×Si − 142.4×$d_g$ − 5.28×$\sigma_g$ − 22.61×$\rho_b$ + 0.46×ID − 2.86×$L$ | | |
| 11 | 148.4 − 0.45×Sa − 1.3×Si − 145.3×$d_g$ − 4.2×$\sigma_g$ − 4.3×$\rho_b$ − 0.053×ID − 0.086×$L$ | | |
| 12 | −896.5 + 9.2×Sa + 9.1×Si + 9.72×Cl − 5.4×dg − 1.37×$\sigma_g$ − 4.7×$\rho_b$ − 0.18×ID − 0.0005×$L$ | | |
| 13 | −76.04 + 0.146×Sa − 0.66×Si − 144.7×$d_g$ − 3.13×$\sigma_g$ − 1.32×$\rho_b$ + 0.03×ID − 0.0606×$L$ | | |
| 14 | −204.9 + 2.17×Sa + 2.15×Si + 2.4×Cl + 2.54×dg − 0.32×$\sigma_g$ − 3.79×$\rho_b$ − 0.2×ID + 0.086×$L$ | | |

Sa: sand (%), Si: silt (%), Cl: clay (%), $\rho_b$: bulk density (g cm$^{-3}$), $d_g$: geometric mean diameter (mm), $\sigma_g$: geometric standard deviation (mm), ARR: average residual reduction. Other CPXR-based pedotransfer functions estimating soil saturated hydraulic conductivity (SHC or $K_{sat}$) from other soil properties are available online at
http://www.knoesis.org/resources/researchers/vahid/behzad.html

Table 4. Mean log-transformed error (MLE) of different models used in this study to estimate $K_{sat}$ for 12 USDA soil texture classes.

| $K_{sat}$ model | USDA soil texture class | | | | | | | | | | | |
|---|---|---|---|---|---|---|---|---|---|---|---|---|
| | Sa[*] | LSa | SaL | L | SiL | Si | SaCL | CL | SiCL | SaC | SiC | C |
| | MLE | | | | | | | | | | | |
| Brakensiek et al. (1984) | 0.18 | 0.59 | 0.76 | -0.09 | -0.74 | -0.73 | 0.49 | -0.77 | -1.03 | -0.32 | -1.87 | -1.08 |
| Campbell and Shiozawa (1994) | -0.82 | -0.66 | -0.70 | -1.80 | -2.63 | -3.11 | -1.10 | -2.35 | -3.23 | -1.96 | -3.97 | -3.11 |
| Cosby et al. (1984) | -0.32 | 0.23 | 0.62 | 0.28 | -0.21 | -0.79 | 0.76 | 0.33 | -0.18 | 0.63 | -0.34 | 0.42 |
| Jabro (1992) | 0.77 | 0.17 | 0.16 | 0.36 | 0.41 | 1.52 | 0.21 | 0.04 | 0.42 | 0.42 | 0.04 | 0.93 |
| Puckett et al. (1985) | -0.29 | -0.06 | -0.13 | -0.49 | -0.36 | -0.11 | -0.75 | -1.41 | -1.48 | -1.80 | -2.75 | -2.85 |
| Dane and Puckett (1994) | 0.05 | 0.41 | 0.51 | 0.27 | 0.32 | 0.37 | 0.14 | -0.34 | -0.45 | -0.59 | -1.38 | -1.30 |
| Saxton et al. (1986) | 0.15 | 0.13 | -0.19 | -0.18 | 0.17 | 1.15 | -0.86 | -1.07 | -0.49 | -1.33 | -1.25 | -0.82 |
| Ghanbarian et al. (2015) | 0.03 | 0.30 | 0.39 | 0.16 | 0.30 | 0.36 | 0.06 | -0.25 | -0.17 | -0.46 | -0.70 | -0.40 |

[*]Sa: Sand, LSa: Loamy sand, SaL: Sandy loam, L: Loam, SiL: Silt loam, Si: Silt, SaCL: Sandy clay loam, CL: Clay loam, SiCL: Silty clay loam, SaC: Sandy clay, SiC: Silty clay, and C: Clay

Table 5. Root mean square log-transformed error (RMSLE) of different models used in this study to estimate $K_{sat}$ for 12 USDA soil texture classes.

| $K_{sat}$ model | USDA soil texture class | | | | | | | | | | | |
|---|---|---|---|---|---|---|---|---|---|---|---|---|
| | Sa[*] | LSa | SaL | L | SiL | Si | SaCL | CL | SiCL | SaC | SiC | C |
| | RMSLE | | | | | | | | | | | |
| Brakensiek et al. (1984) | 0.62 | 0.92 | 1.10 | **0.82**[**] | 1.21 | 1.02 | 1.02 | 1.29 | 1.40 | 1.05 | 2.21 | 1.87 |
| Campbell and Shiozawa (1994) | 0.94 | 0.96 | 1.11 | 2.04 | 2.84 | 3.22 | 1.44 | 2.57 | 3.40 | 2.16 | 4.06 | 3.35 |
| Cosby et al. (1984) | 0.58 | 0.77 | 1.04 | 0.95 | 1.02 | 1.10 | 1.15 | 1.07 | 1.04 | 1.10 | **0.94** | 1.12 |
| Jabro (1992) | 1.10 | 0.93 | 0.95 | 1.02 | 1.14 | 1.71 | 0.98 | 1.16 | 1.10 | 1.12 | 1.18 | 1.59 |
| Puckett et al. (1985) | 0.55 | 0.68 | 0.82 | 1.10 | 1.09 | **0.86** | 1.18 | 1.71 | 1.83 | 2.03 | 2.88 | 3.15 |
| Dane and Puckett (1994) | 0.48 | 0.80 | 0.95 | 0.98 | 1.04 | 0.90 | 0.89 | 1.03 | 1.14 | 1.09 | 1.62 | 1.77 |
| Saxton et al. (1986) | 0.46 | 0.65 | 1.06 | 0.88 | 1.06 | 1.37 | 1.24 | 1.51 | 1.10 | 1.65 | 1.72 | 1.51 |
| Ghanbarian et al. (2015) | **0.42** | **0.64** | **0.77** | 0.82 | **0.83** | 0.87 | **0.70** | **0.74** | **0.80** | **0.80** | 1.01 | **0.82** |

[*]Sa: Sand, LSa: Loamy sand, SaL: Sandy loam, L: Loam, SiL: Silt loam, Si: Silt, SaCL: Sandy clay loam, CL: Clay loam, SiCL: Silty clay loam, SaC: Sandy clay, SiC: Silty clay, and C: Clay
[**]Bold numbers denote the least RMSLE values and most accurate model in each soil texture class

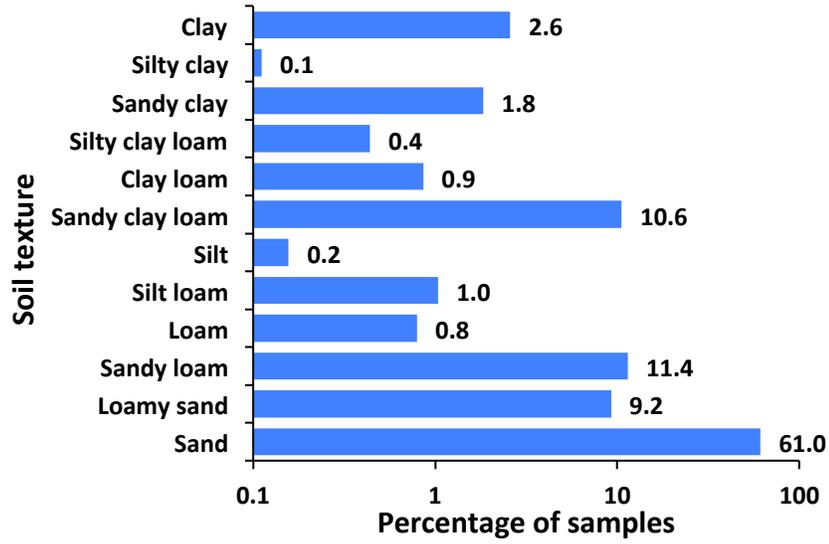

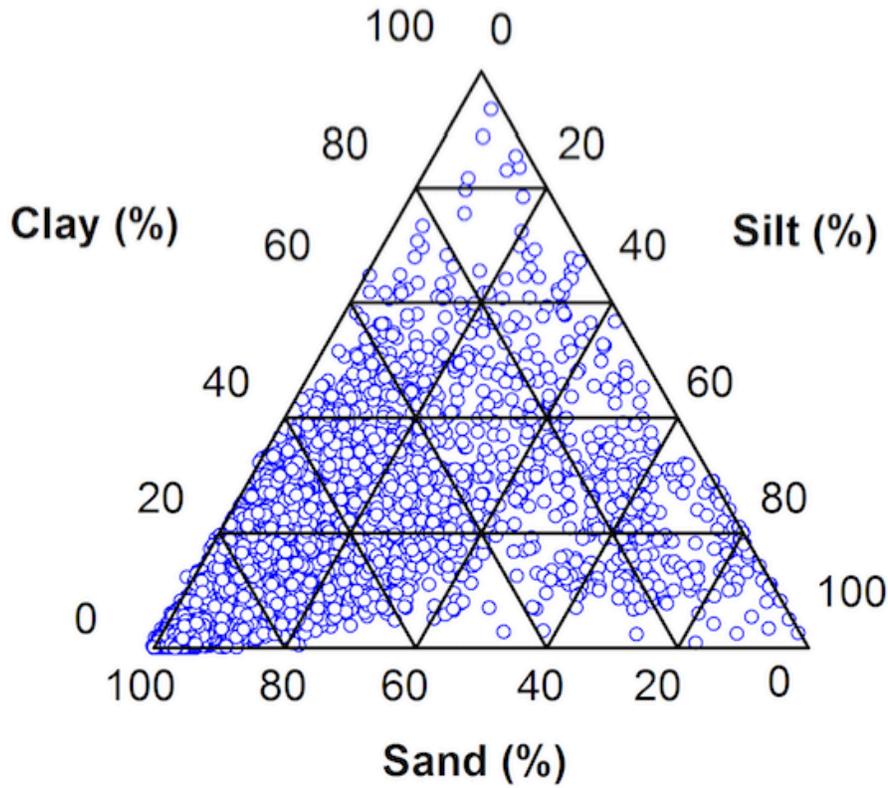

Fig. 1. Percentage of samples in each USDA soil texture class and distribution of samples on the ternary diagram ($n = 19822$).

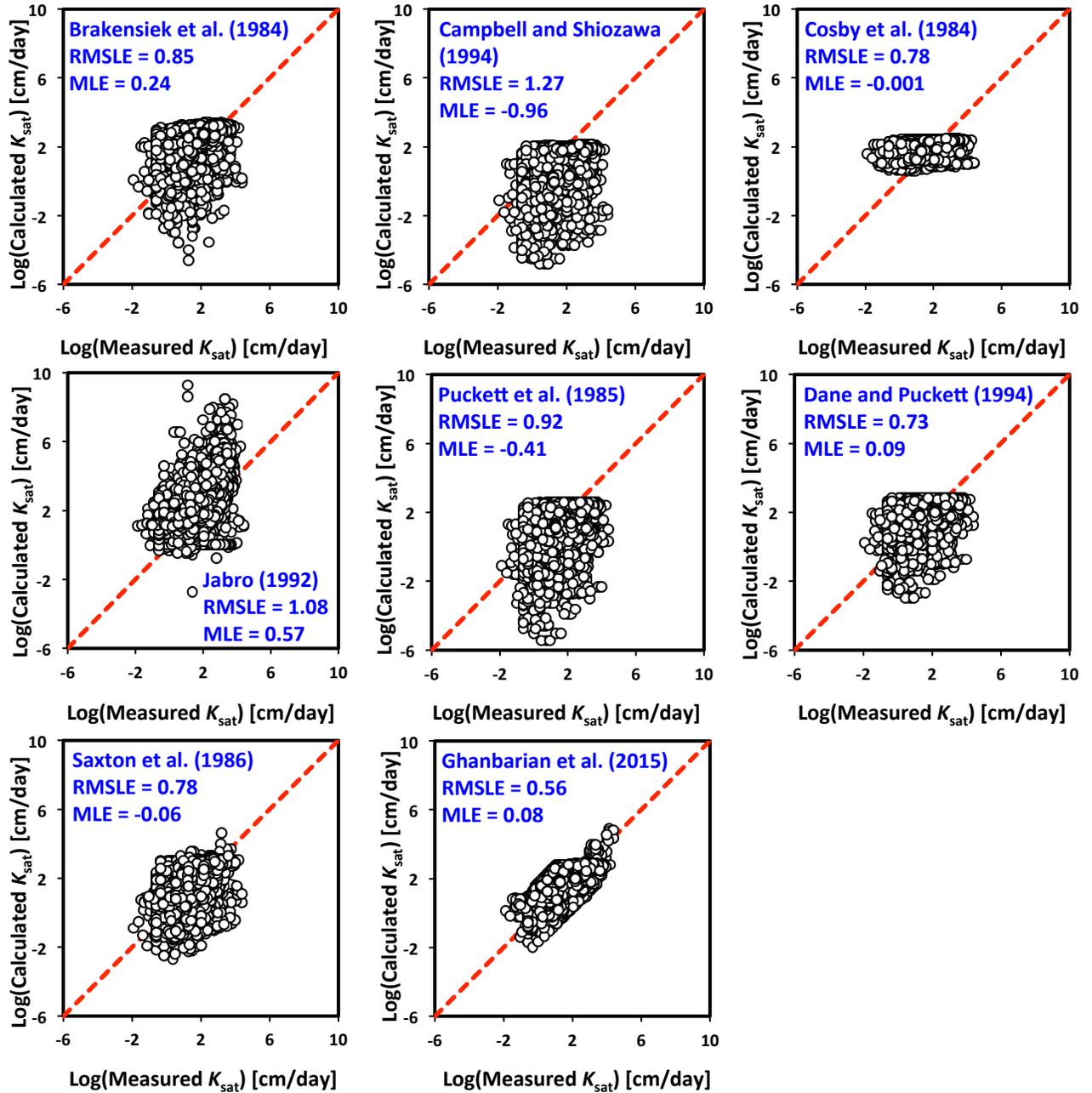

Fig. 2. Comparison of the saturated hydraulic conductivity $K_{sat}$ estimated using eight different methods with the measured ones for all the data used in this study ($n$ = 19822). Note that the number of samples in the elavuation of Jabro (1992) model is 19480 because some samples with either zero silt or clay content were removed from the entire database. The red dashed line represents the 1:1 line.